\documentclass[aps,prl,twocolumn,showpacs,groupedaddress,notitlepage,superscriptaddress,floatfix,longbibliography]{revtex4-1}
\usepackage{amsmath,amsthm,amssymb,amsfonts,enumitem,float,graphicx,physics,hyperref, xcolor}
\usepackage{array,multirow}
\usepackage[normalem]{ulem}
\usepackage{subfigure}
\usepackage[flushleft]{threeparttable}
\usepackage{booktabs}
\usepackage{lipsum}

\theoremstyle{definition}

\newcommand{\PRLsec}[1]{\textit{#1}.---}

\newcommand{\beq}{\begin{equation}}
\newcommand{\eeq}{\end{equation}}
\newcommand{\bes} {\begin{subequations}}
\newcommand{\ees} {\end{subequations}}

\hypersetup{
    colorlinks=true, 
    linkcolor=cyan,
    citecolor=magenta, 
    filecolor=magenta, 
    urlcolor=cyan,
    runcolor=cyan
}

\begin{document}

\title{A deep learning model for noise prediction on near-term quantum devices}

\author{Alexander Zlokapa}
\affiliation{Division of Physics, Mathematics and Astronomy, Caltech, Pasadena, CA 91125, USA}
\author{Alexandru Gheorghiu}
\affiliation{Computing and Mathematical Sciences Department, Caltech, Pasadena, CA 91125, USA}

\begin{abstract}
 We present an approach for a deep-learning compiler of quantum circuits, designed to reduce the output noise of circuits run on a specific device. We train a convolutional neural network on experimental data from a quantum device to learn a hardware-specific noise model. A compiler then uses the trained network as a noise predictor and inserts sequences of gates in circuits so as to minimize expected noise. We tested this approach on the IBM 5-qubit devices and observed a reduction in output noise of $12.3\%$ ($95\%$ CI [11.5\%, 13.0\%]) compared to the circuits obtained by the Qiskit compiler. Moreover, the trained noise model is hardware-specific: applying a noise model trained on one device to another device yields a noise reduction of only $5.2$\% ($95$\% CI [4.9\%, 5.6\%]).  
 These results suggest that device-specific compilers using machine learning may yield higher fidelity operations and provide insights for the design of noise models.
\end{abstract}

\maketitle

\PRLsec{Introduction}%
Quantum computers are expected to offer exponential speedups over classical computers in solving certain computational tasks. The recent demonstration of quantum computational supremacy by the Google group further strengthens the case for the potential of quantum computation~\cite{google}. However, the result also highlighted the limitations of current and near-term quantum devices. It showed that \emph{Noisy Intermediate-Scale Quantum} (NISQ) devices are limited in usability and reliability by errors due to thermal relaxation, measurement, and interactions between adjacent qubits~\cite{preskill2018quantum}.
Mitigating the effect of these errors (or noise), as well as coming up with better noise models are therefore pressing problems of immediate practical relevance. As existing noise models often make simplifying assumptions~\cite{rivas2014quantum, breuer2016colloquium} that fail to capture all of the ways in which errors can corrupt a computation, we are motivated to construct a \emph{learned} noise model using techniques from deep learning~\footnote{Code is available at \url{https://github.com/quantummind/deepQ}.}.

Before protocols for quantum error correction can be implemented, one way to mitigate the effects of noise is to \emph{compile} a given quantum circuit into an equivalent circuit that is less noisy when run on a specific device. By ``less noisy'' we mean that the output distribution of the circuit is closer to the ideal output distribution (obtained when all operations are assumed to be perfect).
Typical approaches for doing this include gate minimization~\cite{gates, Nam2018, ibm-fidelity} and $T$-gate count minimization~\cite{tcount}. In both these cases, however, the function that is being optimized — gate count or $T$-gate count — is merely a proxy for the true objective function, noise. The motivation for these methods is clear: each quantum gate is prone to error and so, by reducing the total number of gates (or the ``noisiest'' gates), the overall number of errors introduced in the circuit is also reduced. However, this paradigm can neglect other effects such as correlated errors between qubits~\cite{harper2019efficient}, rates of different types of errors (such as amplitude damping or dephasing) varying across qubits~\cite{google}, whether the noise is Markovian or not~\cite{rivas2014quantum, breuer2016colloquium} and so on. 

\begin{figure*}
  \centering
  \includegraphics[width=0.9\textwidth]{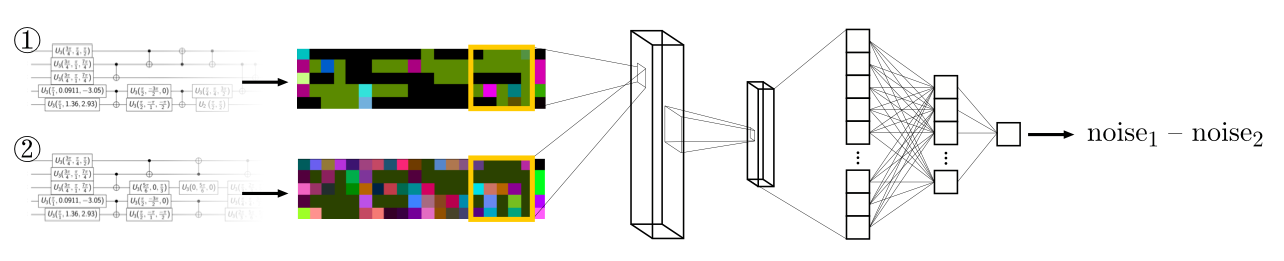}
  \caption{(color online) Schematic representation of how the CNN works. The network learns to predict the difference in output noise between two circuits (assumed to be equivalent). Circuits are encoded as images with each gate being associated a channel (or color). The CNN we used has two convolutional layers and three fully connected layers.}
  \label{fig:cnn}
\end{figure*}

To address some of these issues, we propose using machine learning to \emph{learn a noise model} for a given device. \emph{Convolutional neural networks} (CNNs) can apply convolutional filters on neighboring gates and qubits (see Fig.~\ref{fig:cnn}) and are thus well-suited for recognizing the effects of cross-talk or correlated errors. Due to the ``blackbox'' nature of the network~\footnote{As with most machine-learning applications, understanding how the learned information is represented within the network is challenging. We leave the problem of extracting characteristics of the noise model from the trained network for future work.}, we view it as a \emph{noise predictor} — a function that takes in the description of a circuit and estimates how noisy the circuit's output would be if run on a specific device. This noise predictor can then be used as an objective function for a compiler designed to reduce output noise in quantum circuits without any \emph{a priori} information about the quantum device being used. As our results suggest, this approach can supplement existing techniques for noise mitigation and lead to higher fidelity operations.

\begin{figure}[h]
  \centering
  \includegraphics[width=0.9\columnwidth]{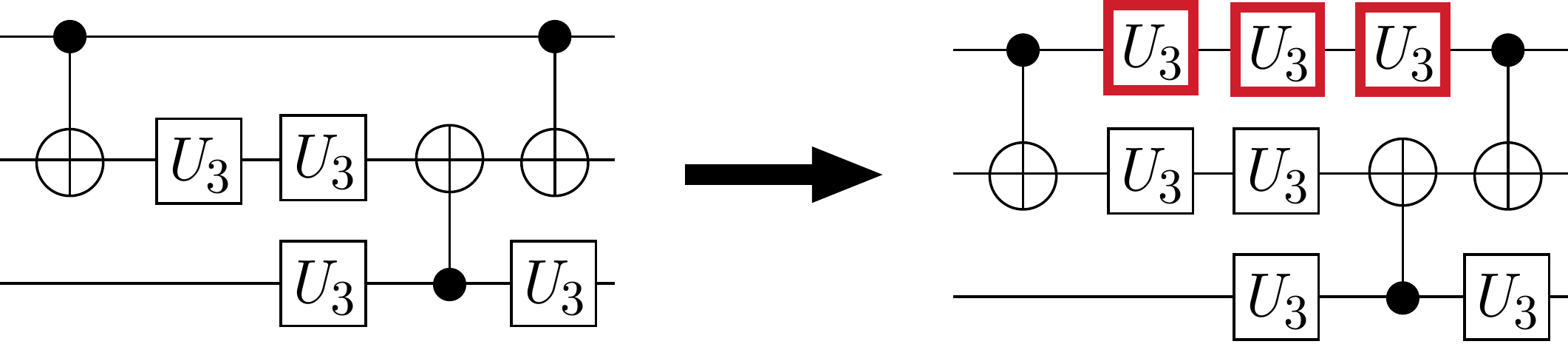}
  \caption{(color online) Gates inserted in the gap of a quantum circuit. A circuit with an idling top qubit (left) is padded with a sequence of gates equivalent to the identity (right), i.e. the gates in red multiply to the identity. Here $U_3$ refers to the parameterized gate shown in Equation~\ref{eqn:u3}.
  }
  \label{fig:compiled}
\end{figure}

To test our approach, we considered the problem of filling gaps in circuits that are run on a device with a restricted topology. Due to the layout of two-qubit gates on a circuit, gaps can often appear as a qubit must ``wait'' for a gate to be applied later in the circuit, as in the left part of Fig.~\ref{fig:compiled}. During this time, these \emph{idling qubits} undergo free evolution, under some Hamiltonian $H$, and their state can drift.  
However, through the application of certain sequences of gates (that collectively act as the identity), this drift can be averaged out so that its net effect is suppressed. 
Of course, these sequences are not something we would arrive at through gate minimization since filling these gaps with unitaries increases total gate count. Instead, we can train a CNN to predict the effect of adding various sequences of gates within these gaps. We then use this to identify those sequences that lead to a net reduction in output noise.

This specific approach of adding gates in a circuit so as to reduce noise and suppress coherent non-Markovian effects is related to two concepts from the literature.
The first is \emph{dynamical decoupling}~\cite{viola1998dynamical,duan1999suppressing,zanardi1999symmetrizing,viola1999dynamical,lidar2014review}, which reduces the interaction between a system and its environment by applying repeated unitaries (pulses) that are equivalent to the identity.
Specifically, instead of letting a qubit evolve under the action of $e^{-iHt}$, for some Hamiltonian $H$, the pulse $P$ is applied so that the state is acted upon by $P e^{-iHt}$. The state is then left to evolve again for a time $t$ and another pulse $Q$ is applied. If $P$ and $Q$ are such that $Q H P = -H$, we can see that the effect of two such pulses is to cancel out the action of $H$:
\begin{equation*}
    Q e^{-iHt} P e^{-iHt} = e^{iHt} e^{-iHt} = \mathbb{Id} 
\end{equation*}
Thus, adding these so-called \emph{DD sequences} in the circuit can reduce output noise. Note that if $P$ and $Q$ are the only pulses to act within a gap, it must be that $P = Q^{\dagger}$, so that the combined action is equivalent to the identity. 

The second related concept is that of \emph{randomized compiling}~\cite{wallman2016noise, erhard2019characterizing}, where random unitaries (that collectively act as identity) are added to a circuit so as to turn \emph{coherent errors} into \emph{stochastic Pauli errors}. Since coherent errors tend to be more detrimental to quantum computations, this has the effect of reducing output noise. In some cases the random unitaries are not added into the circuit itself, but are instead ``absorbed'' into neighbouring gates and the resulting equivalent circuit is run on the device.
In our approach, while in the training of our CNN we consider random sequences of gates to insert into the gaps, the compiler will use the trained CNN to search for \emph{optimal sequences} (i.e. those sequences that lead to the greatest reduction in output noise, as predicted by the CNN). In that sense, our compiler is not ``randomized''.

\PRLsec{Deep learning compiler}%
At a high level, our approach consists of first creating the learned noise model and then using it as an objective function to be optimized when doing circuit compilation.
The noise model (or noise predictor) is a \emph{convolutional neural network} (CNN) that is trained to predict the \emph{difference in noise} between two equivalent circuits evaluated on a quantum device (Fig.~\ref{fig:cnn}).
The network takes in \emph{pairs} of (equivalent) circuits encoded as images (each pixel representing a gate). To ensure these encodings are consistent, we topologically sort the graph representation of each circuit before encoding it as a multi-channel image (one channel for each of 4 native gate types). The specific CNN we considered has two convolutional layers ($5\times 5$ and $3\times 3$ filters) and three fully-connected layers and is trained with a step decay schedule of learning rates and an optimized batch size for stable training. Training examples consist of pairs of circuits together with the difference in output noise when the circuits are run on a specific device.

\begin{figure*}[ht]
  \centering
  \includegraphics[width=0.9\textwidth]{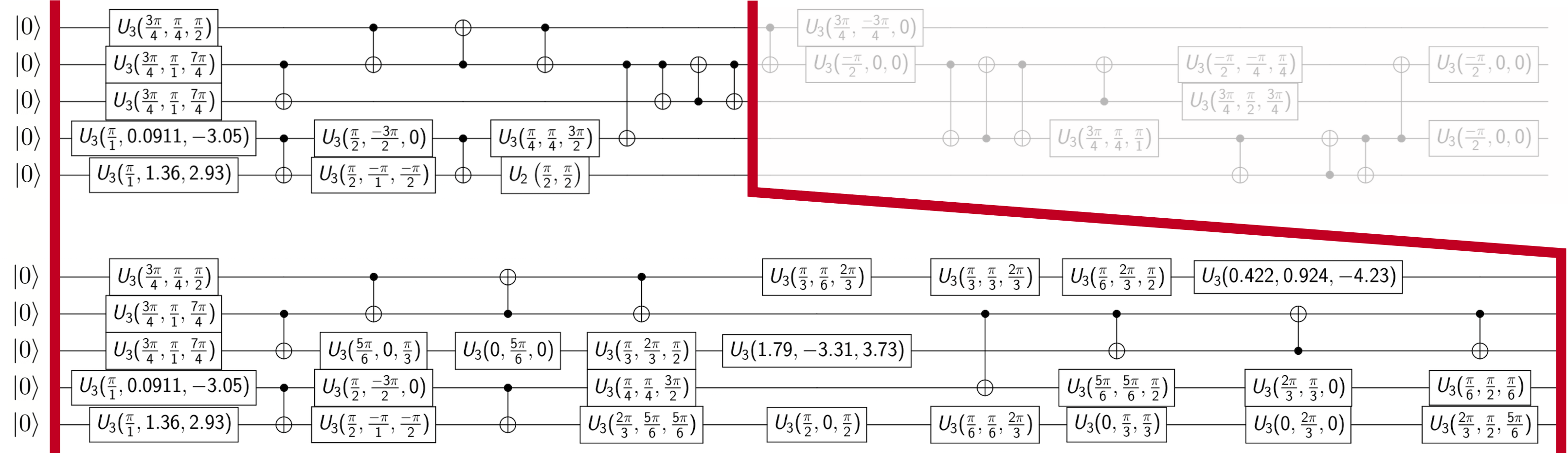}
  \caption{(color online) Circuit compiled by Qiskit (top) and by Qiskit together with the deep learning compiler for the Burlington device (bottom). The top circuit is the complete $UU^\dagger$ circuit, whereas only a portion is shown for the bottom one. (Gap artifacts are visible in the bottom circuit because gates are scheduled to be applied as late as possible, causing gates to be shifted to the right when run on hardware.) Both circuits have depth 20. Top: 35 gates, fraction of ones in the output is 0.32. Bottom: 80 gates, fraction of ones in the output is 0.27. Full circuits available at \url{https://github.com/quantummind/deepQ}.}
  \label{fig:compiled_large}
\end{figure*}

Once the network has been trained, we can perform circuit compilation. Starting from an initial circuit $C_0$, we generate a set of equivalent circuits $\mathcal{S} = \{C_1, \dots, C_m\}$. Each $C_i \in \mathcal{S}$ corresponds to the same unitary operation as $C_0$, even though the circuit representation is different. Creating the equivalent circuits is done by inserting sequences of random gates that map to the identity into the gaps of $C_0$. We refer to these as \emph{random gate sequences}. The deep learning noise model is then used to perform pairwise tournament comparisons in expected noise between $m=1000$ circuits from $\mathcal{S}$, with the goal of finding the least noisy circuit, $C_i$.
To benchmark this approach, we run both $C_i$ and $C_0$ repeatedly on a quantum device and estimate the difference in output noise between the circuits.

Quantifying the output noise of a circuit $C$ relative to a device can be done in a number of ways. One approach is to compute the total variation distance between the ideal output distribution of $C$ (for some fixed input, say $\ket{00...0}$) and the output distribution obtained from running the circuit.
For general circuits this would scale exponentially in the number of qubits, so we instead consider only circuits of the form $UU^\dagger$, for some unitary $U$. For these circuits, when acting on $\ket{00...0}$, we expect to see an all-zero output (after measuring in the computational basis). We will refer to this as the \emph{singleton distribution}. 
The total variation distance between the singleton distribution and some other distribution, $p : \{0, 1\}^n \to [0, 1]$, is simply $1 - p(0^n)$. This, however, would not be a good measure for output noise: suppose we have the distributions $p(0^n) = 1/2$, $p(10^{n-1}) = 1/2$ and $q(0^n) = 1/2$, $q(1^n) = 1/2$. Both of these have total variation distance $1/2$ from the singleton. However, a circuit that outputs $q$ is expected to be noisier than one that outputs $p$. For the latter, a single bit of the output is being flipped with probability $1/2$, whereas for the former all bits are being flipped with probability $1/2$.
As such, our measure of noise will be the average \emph{Hamming weight} which we estimate by counting the number of ones in the output.

The $U$ circuits we generated are similar to those used in the recent Google quantum supremacy experiment~\cite{google}. These comprise of multiple \emph{cycles}, each cycle consisting of a layer of single-qubit gates and one two-qubit gate. The single-qubit gates are selected randomly from $\{\sqrt{X}, \sqrt{W}, \sqrt{Z}\}$, where $W = (X+Y)/\sqrt{2}$; the two-qubit gate is a $CX$ gate between an \emph{arbitrary} pair of qubits.

We generated a dataset of 1,000 random 5-qubit circuits each with 5 cycles to create highly entangled states.
Of these 1,000 circuits, 800 are used for training the CNN, 100 are used for validation and 100 for testing.
If we denote one of these circuits as $U$, we first create $C = UU^{\dagger}$ and compile it to full optimization with the IBM Qiskit compiler (\texttt{optimization\_level=3}) for the device corresponding to the noise model. This yields a circuit expected to have the lowest possible noise by standard compilation techniques, both by relabeling qubits to apply fewer operations to noisier couplings and by rewriting the circuit into an approximate form that is expected to result in a state closer to the ideal state. The gates in the circuit are expressed in the native IBM Q gate basis~\cite{qiskit} $\{U_1, U_2, U_3, CX\}$,
where the parametrized single-qubit $U_3$ gate is:
 \begin{equation} \label{eqn:u3}
 U_3(\theta, \phi, \lambda) = \left( \begin{matrix}
 \cos(\theta / 2) & -e^{i \lambda} \sin(\theta/2) \\
 e^{i \phi} \sin(\theta / 2) & e^{i(\lambda + \phi)} \cos(\theta/2) \\
 \end{matrix} \right),
 \end{equation}
 and $U_1(\lambda) = U_3(0, 0, \lambda), U_2(\phi, \lambda) = U_3(\pi/2, \phi, \lambda)$.
It should be noted that due to the limited connectivity between qubits, the depth of the circuit typically increases as additional two-qubit gates are required. So while the original circuit has depth 20 ($C$ has twice the depth of $U$ and $U$ has 5 cycles, each of depth 2), after compilation, the circuit depth of $UU^\dagger$ ranges from 20 to 44 with an average depth of 26.3.

For each of these 1,000 circuits obtained after compiling with Qiskit, we generate 15 equivalent circuits by inserting random gate sequences in their gaps (Fig.~\ref{fig:compiled}).
The sequences are constructed with $U_3$ gates. To create a sequence of length $l$, we first make $(l-1)$ $U_3$ gates with the parameters $\theta$, $\phi$ and $\lambda$ chosen uniformly at random from $\{0, \pi/6, ... \; 11\pi/6 \}$. We then compute the inverse of this $(l-1)$-length sequence as the last $U_3$ gate.

These final circuits are then used for training and testing. The neural network is trained with $800 \times 16$ circuits. Each circuit is run on an IBM 5-qubit device 5,000 times and the circuit's noise is taken to be the average number of ones measured in the output. The training examples for the network then consist of pairs of equivalent circuits from the 16 corresponding to each base circuit. Early stopping on the validation set is used to prevent overfitting. We also enforce symmetry on the network during prediction (so that $[\mathrm{noise}(C_1) - \mathrm{noise}(C_2)]$ is always $-[\mathrm{noise}(C_2) - \mathrm{noise}(C_1)]$).
The network will thus learn a noise model for the given IBM device.

\PRLsec{Results on IBM Q 5-qubit devices}%
The experimental results we obtained from multiple IBM Q 5-qubit devices for our compiler are summarized in Table~\ref{tab:perf} with sample circuits shown in Fig.~\ref{fig:compiled_large}. We demonstrate learned noise models for two devices: IBM Q Burlington and IBM Q London. The circuits compiled by Qiskit alone and by Qiskit together with the deep learning compiler are evaluated on \emph{all} available 5-qubit IBM devices, not just the device corresponding to the trained noise model. To check that our approach is indeed finding close-to-optimal gate sequences, we also include a benchmark of the application of random gates in circuit gaps.

We find that the compiler augmented with deep learning performs significantly better than the Qiskit compiler alone. For the device on which the noise model was learned, noise is reduced on average by $11\%$ on Burlington and $13\%$ on London using deep learning compared to using only Qiskit. On the other devices, the reduction is only around $5$\%. This discrepancy seems to suggest that a device-specific noise model was indeed learned. Significant improvement is also observed compared to just adding random sequences of unitaries in the circuit gaps. For these random sequences, the improvement over gapped circuits is around $4.5\%$ for Burlington and $7.3\%$ for London, demonstrating that the compiler optimized the selection of gate sequences. Compiled circuits for which all inserted gate sequence lengths were multiples of 4 were also benchmarked against the standard $XYXY$ sequence used in dynamical decoupling~\cite{viola1999dynamical}. Circuits with $XYXY$ sequences were found to have an average of $6.5\%$ (95\% CI [2.1\%, 10.6\%]) more noise than the deep learning sequences across the Burlington and London devices.

\begin{table}[h]
  \caption{Percent noise improvement of the deep learning noise model compiler compared to the IBM Qiskit compiler. Boldface row indicates the device on which the noise model was learned. Rows marked ``random'' have random $U_3$ gate sequences (equivalent to the identity) in circuit gaps. Yorktown has a bowtie architecture; all other devices have T-shaped architectures. We report 95\% confidence intervals.}
  \label{tab:perf}
  \begin{ruledtabular}
  \begin{tabular}{lrr}
    \toprule
    Device&$\Delta$ Noise (Burl.)&$\Delta$ Noise (Lond.)\\
    \hline\\[-2.0ex]
    Burlington & \textbf{11\% [10\%, 12\%]} & 10\% [8\%, 11\%]\\
    Burl. random & 4.5\% [4.4\%, 4.6\%] & 4.5\% [4.4\%, 4.6\%]\\
    Essex & 4\% [3\%, 5\%] & 4\% [2\%, 5\%]\\
    London & 5\% [4\%, 6\%] & \textbf{13\% [13\%, 14\%]}\\
    Lond. random & 7.3\% [7.2\%, 7.5\%] & 7.3\% [7.2\%, 7.5\%]\\
    Ourense & 6\% [5\%, 7\%] & 7\% [6\%, 8\%]\\
    Vigo & 4\% [4\%, 5\%] & 3\% [2\%, 4\%]\\
    Yorktown & 6\% [5\%, 7\%] & 4\% [3\%, 5\%]\\
    \bottomrule
  \end{tabular}
  \end{ruledtabular}
\end{table}

\begin{figure}[h]
  \centering
  \hspace{-7mm}%
  \includegraphics[width=0.95\columnwidth]{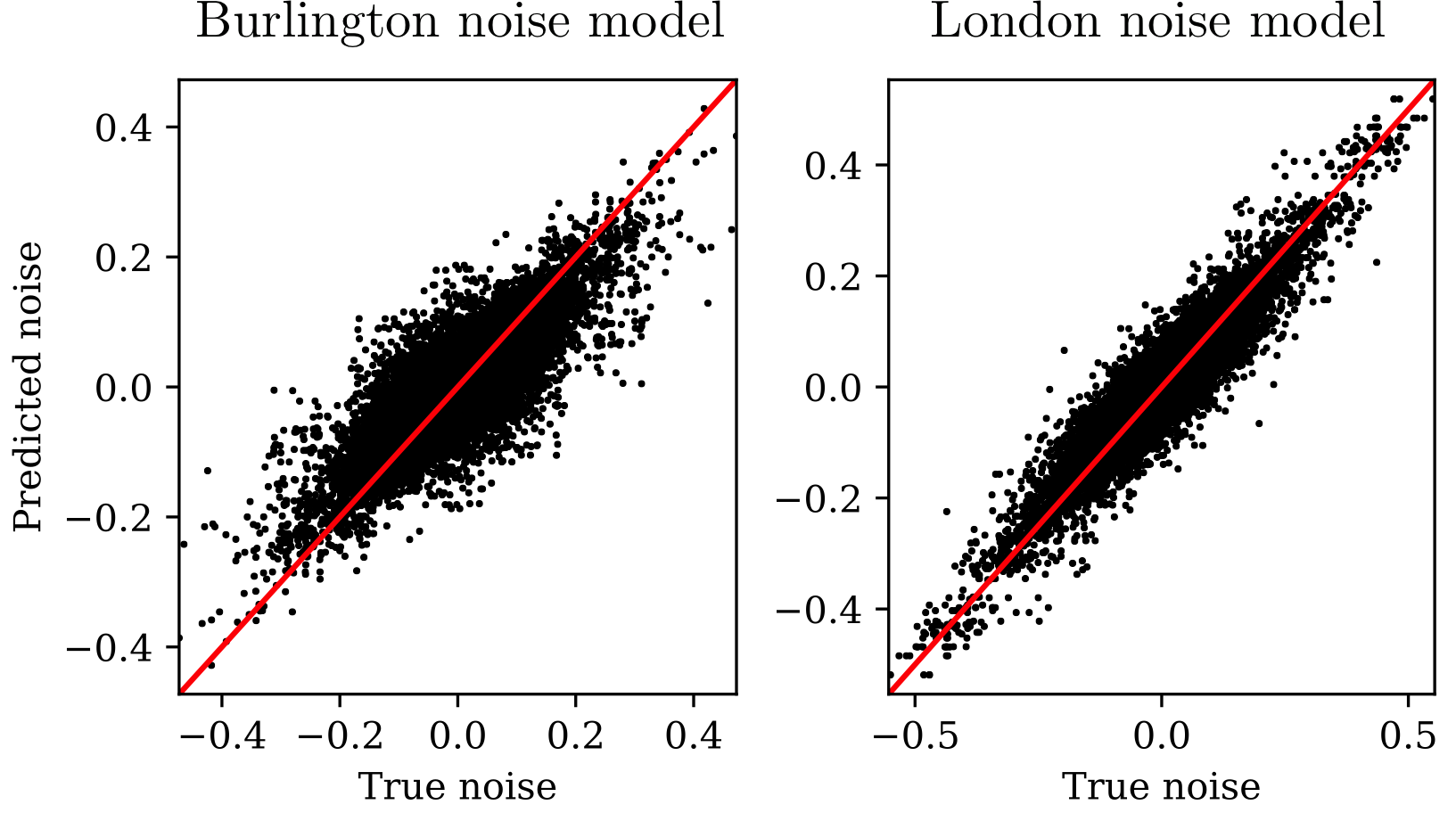}%
  \caption{(color online) Predicted and true noise of 100 random circuits in the test dataset for the Burlington noise model ($R^2 = 0.66$) and London noise model ($R^2=0.86$). An ideal predictor would yield the red line.}
  \label{fig:noise}
\end{figure}

We also evaluate the performance of the deep learning model on its own. As mentioned, the CNN is trained to predict output noise and we can use our test set of $100 \times 16$ random circuits to see how well it performed in that task. 
The results are shown in Fig.~\ref{fig:noise}. For the two IBM devices, we find that the unexplained variance in output noise is $34\%$ and $14\%$ respectively, i.e. $R^2 = 0.66$ and $0.86$.
We emphasize that the network achieved this performance from the trained data alone, without any built-in knowledge of the gate fidelities or coherence times of the quantum device.
The trained network can thus be used to rate the performance of gate sequences without having to run the circuits on the device.

\PRLsec{Discussion}%
Our results suggest that using deep learning to learn a hardware-specific noise model can lead to circuit compilers that improve the fidelity of quantum circuits running on noisy devices. Specifically, we found that on IBM devices our approach leads to an average reduction of 12.3\% in output noise when compared to the circuits produced by the Qiskit compiler. Moreover, our results indicate that the learned noise model is specific to the device on which it was trained. 
This is the case even though no \emph{a priori} knowledge of gate fidelities, qubit coherence times or any other hardware-specific information, was provided to the model. The network learned to predict noise from examples of circuits that ran on the device and the estimated noise in their output.

While we focused on the specific case of inserting sequences of gates into circuit gaps, one could ask whether this approach can be used to predict noise for general circuits and use more sophisticated compilers that search for more elaborate equivalent circuits. The difficulty in pursuing such an approach is the \emph{curse of dimensionality} associated with having fewer constraints. The space of equivalent circuits that could be explored would increase significantly
requiring additional techniques in machine learning to efficiently find an optimal compilation.
Nevertheless, the fact that our approach learned device-specific noise suggests that this methodology can be useful in a broader context.

We believe our approach opens up several new lines of investigation. Although the CNN is currently treated as a blackbox, it would be useful to understand how effects such as cross-talk and non-Markovianity are represented within the network. It would be interesting to see whether other deep learning approaches that use convolutional networks as building blocks, such as deep reinforcement learning, are useful for noise reduction. Finally, while we restricted the network to adding sequences of gates in a known way (similar to randomized compiling or dynamical decoupling), we would like to see whether the general machine learning approach can provide new techniques for noise mitigation.

\acknowledgements
The authors thank Thomas Vidick, Daniel Lidar, John Preskill and Matty Hoban for helpful discussions and comments.
AZ is partially supported by Caltech's Intelligent Quantum Networks and Technologies (INQNET) research program and by the DOE/HEP
QuantISED program grant, Quantum Machine Learning
and Quantum Computation Frameworks (QMLQCF) for
HEP, award number DE-SC0019227.
AG is supported by MURI Grant
FA9550-18-1-0161 and the IQIM, an NSF Physics Frontiers Center (NSF Grant PHY-1125565)
with support of the Gordon and Betty Moore Foundation (GBMF-12500028).

\emph{Note added} — Shortly after the completion of this work, the independent works of Strikis et al.~\cite{strikis2020learning} and Czarnik et al.~\cite{czarnik2020error} appeared on the arXiv. These works also propose approaches for quantum error mitigation using learned noise models. 

\bibliography{references}

\begin{thebibliography}{21}%
\makeatletter
\providecommand \@ifxundefined [1]{%
 \@ifx{#1\undefined}
}%
\providecommand \@ifnum [1]{%
 \ifnum #1\expandafter \@firstoftwo
 \else \expandafter \@secondoftwo
 \fi
}%
\providecommand \@ifx [1]{%
 \ifx #1\expandafter \@firstoftwo
 \else \expandafter \@secondoftwo
 \fi
}%
\providecommand \natexlab [1]{#1}%
\providecommand \enquote  [1]{``#1''}%
\providecommand \bibnamefont  [1]{#1}%
\providecommand \bibfnamefont [1]{#1}%
\providecommand \citenamefont [1]{#1}%
\providecommand \href@noop [0]{\@secondoftwo}%
\providecommand \href [0]{\begingroup \@sanitize@url \@href}%
\providecommand \@href[1]{\@@startlink{#1}\@@href}%
\providecommand \@@href[1]{\endgroup#1\@@endlink}%
\providecommand \@sanitize@url [0]{\catcode `\\12\catcode `\$12\catcode
  `\&12\catcode `\#12\catcode `\^12\catcode `\_12\catcode `\%12\relax}%
\providecommand \@@startlink[1]{}%
\providecommand \@@endlink[0]{}%
\providecommand \url  [0]{\begingroup\@sanitize@url \@url }%
\providecommand \@url [1]{\endgroup\@href {#1}{\urlprefix }}%
\providecommand \urlprefix  [0]{URL }%
\providecommand \Eprint [0]{\href }%
\providecommand \doibase [0]{http://dx.doi.org/}%
\providecommand \selectlanguage [0]{\@gobble}%
\providecommand \bibinfo  [0]{\@secondoftwo}%
\providecommand \bibfield  [0]{\@secondoftwo}%
\providecommand \translation [1]{[#1]}%
\providecommand \BibitemOpen [0]{}%
\providecommand \bibitemStop [0]{}%
\providecommand \bibitemNoStop [0]{.\EOS\space}%
\providecommand \EOS [0]{\spacefactor3000\relax}%
\providecommand \BibitemShut  [1]{\csname bibitem#1\endcsname}%
\let\auto@bib@innerbib\@empty
\bibitem [{\citenamefont {Arute}\ \emph {et~al.}(2019)\citenamefont {Arute},
  \citenamefont {Arya}, \citenamefont {Babbush}, \citenamefont {Bacon},
  \citenamefont {Bardin}, \citenamefont {Barends}, \citenamefont {Biswas},
  \citenamefont {Boixo}, \citenamefont {Brandao}, \citenamefont {Buell} \emph
  {et~al.}}]{google}%
  \BibitemOpen
  \bibfield  {author} {\bibinfo {author} {\bibfnamefont {Frank}\ \bibnamefont
  {Arute}}, \bibinfo {author} {\bibfnamefont {Kunal}\ \bibnamefont {Arya}},
  \bibinfo {author} {\bibfnamefont {Ryan}\ \bibnamefont {Babbush}}, \bibinfo
  {author} {\bibfnamefont {Dave}\ \bibnamefont {Bacon}}, \bibinfo {author}
  {\bibfnamefont {Joseph~C}\ \bibnamefont {Bardin}}, \bibinfo {author}
  {\bibfnamefont {Rami}\ \bibnamefont {Barends}}, \bibinfo {author}
  {\bibfnamefont {Rupak}\ \bibnamefont {Biswas}}, \bibinfo {author}
  {\bibfnamefont {Sergio}\ \bibnamefont {Boixo}}, \bibinfo {author}
  {\bibfnamefont {Fernando~GSL}\ \bibnamefont {Brandao}}, \bibinfo {author}
  {\bibfnamefont {David~A}\ \bibnamefont {Buell}},  \emph {et~al.},\ }\bibfield
   {title} {\enquote {\bibinfo {title} {Quantum supremacy using a programmable
  superconducting processor},}\ }\href {\doibase 10.1038/s41586-019-1666-5}
  {\bibfield  {journal} {\bibinfo  {journal} {Nature}\ }\textbf {\bibinfo
  {volume} {574}},\ \bibinfo {pages} {505--510} (\bibinfo {year}
  {2019})}\BibitemShut {NoStop}%
\bibitem [{\citenamefont {Preskill}(2018)}]{preskill2018quantum}%
  \BibitemOpen
  \bibfield  {author} {\bibinfo {author} {\bibfnamefont {John}\ \bibnamefont
  {Preskill}},\ }\bibfield  {title} {\enquote {\bibinfo {title} {Quantum
  {C}omputing in the {NISQ} era and beyond},}\ }\href {\doibase
  10.22331/q-2018-08-06-79} {\bibfield  {journal} {\bibinfo  {journal}
  {{Quantum}}\ }\textbf {\bibinfo {volume} {2}},\ \bibinfo {pages} {79}
  (\bibinfo {year} {2018})}\BibitemShut {NoStop}%
\bibitem [{\citenamefont {Rivas}\ \emph {et~al.}(2014)\citenamefont {Rivas},
  \citenamefont {Huelga},\ and\ \citenamefont {Plenio}}]{rivas2014quantum}%
  \BibitemOpen
  \bibfield  {author} {\bibinfo {author} {\bibfnamefont {{\'{A}}ngel}\
  \bibnamefont {Rivas}}, \bibinfo {author} {\bibfnamefont {Susana~F}\
  \bibnamefont {Huelga}}, \ and\ \bibinfo {author} {\bibfnamefont {Martin~B}\
  \bibnamefont {Plenio}},\ }\bibfield  {title} {\enquote {\bibinfo {title}
  {Quantum non-markovianity: characterization, quantification and detection},}\
  }\href {\doibase 10.1088/0034-4885/77/9/094001} {\bibfield  {journal}
  {\bibinfo  {journal} {Reports on Progress in Physics}\ }\textbf {\bibinfo
  {volume} {77}},\ \bibinfo {pages} {094001} (\bibinfo {year}
  {2014})}\BibitemShut {NoStop}%
\bibitem [{\citenamefont {Breuer}\ \emph {et~al.}(2016)\citenamefont {Breuer},
  \citenamefont {Laine}, \citenamefont {Piilo},\ and\ \citenamefont
  {Vacchini}}]{breuer2016colloquium}%
  \BibitemOpen
  \bibfield  {author} {\bibinfo {author} {\bibfnamefont {Heinz-Peter}\
  \bibnamefont {Breuer}}, \bibinfo {author} {\bibfnamefont {Elsi-Mari}\
  \bibnamefont {Laine}}, \bibinfo {author} {\bibfnamefont {Jyrki}\ \bibnamefont
  {Piilo}}, \ and\ \bibinfo {author} {\bibfnamefont {Bassano}\ \bibnamefont
  {Vacchini}},\ }\bibfield  {title} {\enquote {\bibinfo {title} {Colloquium:
  Non-markovian dynamics in open quantum systems},}\ }\href {\doibase
  10.1103/RevModPhys.88.021002} {\bibfield  {journal} {\bibinfo  {journal}
  {Rev. Mod. Phys.}\ }\textbf {\bibinfo {volume} {88}},\ \bibinfo {pages}
  {021002} (\bibinfo {year} {2016})}\BibitemShut {NoStop}%
\bibitem [{Note1()}]{Note1}%
  \BibitemOpen
  \bibinfo {note} {Code is available at \protect \url
  {https://github.com/quantummind/deepQ}.}\BibitemShut {Stop}%
\bibitem [{\citenamefont {Kliuchnikov}\ and\ \citenamefont
  {Maslov}(2013)}]{gates}%
  \BibitemOpen
  \bibfield  {author} {\bibinfo {author} {\bibfnamefont {Vadym}\ \bibnamefont
  {Kliuchnikov}}\ and\ \bibinfo {author} {\bibfnamefont {Dmitri}\ \bibnamefont
  {Maslov}},\ }\bibfield  {title} {\enquote {\bibinfo {title} {Optimization of
  {C}lifford circuits},}\ }\href {\doibase 10.1103/PhysRevA.88.052307}
  {\bibfield  {journal} {\bibinfo  {journal} {Phys. Rev. A}\ }\textbf {\bibinfo
  {volume} {88}},\ \bibinfo {pages} {052307} (\bibinfo {year}
  {2013})}\BibitemShut {NoStop}%
\bibitem [{\citenamefont {Nam}\ \emph {et~al.}(2018)\citenamefont {Nam},
  \citenamefont {Ross}, \citenamefont {Su}, \citenamefont {Childs},\ and\
  \citenamefont {Maslov}}]{Nam2018}%
  \BibitemOpen
  \bibfield  {author} {\bibinfo {author} {\bibfnamefont {Yunseong}\
  \bibnamefont {Nam}}, \bibinfo {author} {\bibfnamefont {Neil~J.}\ \bibnamefont
  {Ross}}, \bibinfo {author} {\bibfnamefont {Yuan}\ \bibnamefont {Su}},
  \bibinfo {author} {\bibfnamefont {Andrew~M.}\ \bibnamefont {Childs}}, \ and\
  \bibinfo {author} {\bibfnamefont {Dmitri}\ \bibnamefont {Maslov}},\
  }\bibfield  {title} {\enquote {\bibinfo {title} {Automated optimization of
  large quantum circuits with continuous parameters},}\ }\href {\doibase
  10.1038/s41534-018-0072-4} {\bibfield  {journal} {\bibinfo  {journal} {npj
  Quantum Information}\ }\textbf {\bibinfo {volume} {4}},\ \bibinfo {pages}
  {23} (\bibinfo {year} {2018})}\BibitemShut {NoStop}%
\bibitem [{\citenamefont {Dueck}\ \emph {et~al.}(2018)\citenamefont {Dueck},
  \citenamefont {Pathak}, \citenamefont {Rahman}, \citenamefont {Shukla},\ and\
  \citenamefont {Banerjee}}]{ibm-fidelity}%
  \BibitemOpen
  \bibfield  {author} {\bibinfo {author} {\bibfnamefont {Gerhard~W.}\
  \bibnamefont {Dueck}}, \bibinfo {author} {\bibfnamefont {Anirban}\
  \bibnamefont {Pathak}}, \bibinfo {author} {\bibfnamefont {Md.~Mazder}\
  \bibnamefont {Rahman}}, \bibinfo {author} {\bibfnamefont {Abhishek}\
  \bibnamefont {Shukla}}, \ and\ \bibinfo {author} {\bibfnamefont {Anindita}\
  \bibnamefont {Banerjee}},\ }\bibfield  {title} {\enquote {\bibinfo {title}
  {Optimization of circuits for {IBM}'s five-qubit quantum computers},}\
  }\href@noop {} {\bibfield  {journal} {\bibinfo  {journal} {2018 21st
  Euromicro Conference on Digital System Design (DSD)}\ ,\ \bibinfo {pages}
  {680--684}} (\bibinfo {year} {2018})}\BibitemShut {NoStop}%
\bibitem [{\citenamefont {Abdessaied}\ \emph {et~al.}(2014)\citenamefont
  {Abdessaied}, \citenamefont {Soeken},\ and\ \citenamefont
  {Drechsler}}]{tcount}%
  \BibitemOpen
  \bibfield  {author} {\bibinfo {author} {\bibfnamefont {Nabila}\ \bibnamefont
  {Abdessaied}}, \bibinfo {author} {\bibfnamefont {Mathias}\ \bibnamefont
  {Soeken}}, \ and\ \bibinfo {author} {\bibfnamefont {Rolf}\ \bibnamefont
  {Drechsler}},\ }\bibfield  {title} {\enquote {\bibinfo {title} {Quantum
  circuit optimization by {H}adamard gate reduction},}\ }in\ \href@noop {}
  {\emph {\bibinfo {booktitle} {Reversible Computation}}},\ \bibinfo {editor}
  {edited by\ \bibinfo {editor} {\bibfnamefont {Shigeru}\ \bibnamefont
  {Yamashita}}\ and\ \bibinfo {editor} {\bibfnamefont {Shin-ichi}\ \bibnamefont
  {Minato}}}\ (\bibinfo  {publisher} {Springer International Publishing},\
  \bibinfo {address} {Cham},\ \bibinfo {year} {2014})\ pp.\ \bibinfo {pages}
  {149--162}\BibitemShut {NoStop}%
\bibitem [{\citenamefont {Harper}\ \emph {et~al.}(2019)\citenamefont {Harper},
  \citenamefont {Flammia},\ and\ \citenamefont
  {Wallman}}]{harper2019efficient}%
  \BibitemOpen
  \bibfield  {author} {\bibinfo {author} {\bibfnamefont {Robin}\ \bibnamefont
  {Harper}}, \bibinfo {author} {\bibfnamefont {Steven~T.}\ \bibnamefont
  {Flammia}}, \ and\ \bibinfo {author} {\bibfnamefont {Joel~J.}\ \bibnamefont
  {Wallman}},\ }\href@noop {} {\enquote {\bibinfo {title} {Efficient learning
  of quantum noise},}\ } (\bibinfo {year} {2019}),\ \Eprint
  {http://arxiv.org/abs/1907.13022} {arXiv:1907.13022 [quant-ph]} \BibitemShut
  {NoStop}%
\bibitem [{Note2()}]{Note2}%
  \BibitemOpen
  \bibinfo {note} {As with most machine-learning applications, understanding
  how the learned information is represented within the network is challenging.
  We leave the problem of extracting characteristics of the noise model from
  the trained network for future work.}\BibitemShut {Stop}%
\bibitem [{\citenamefont {Viola}\ and\ \citenamefont
  {Lloyd}(1998)}]{viola1998dynamical}%
  \BibitemOpen
  \bibfield  {author} {\bibinfo {author} {\bibfnamefont {Lorenza}\ \bibnamefont
  {Viola}}\ and\ \bibinfo {author} {\bibfnamefont {Seth}\ \bibnamefont
  {Lloyd}},\ }\bibfield  {title} {\enquote {\bibinfo {title} {Dynamical
  suppression of decoherence in two-state quantum systems},}\ }\href {\doibase
  10.1103/PhysRevA.58.2733} {\bibfield  {journal} {\bibinfo  {journal} {Phys.
  Rev. A}\ }\textbf {\bibinfo {volume} {58}},\ \bibinfo {pages} {2733--2744}
  (\bibinfo {year} {1998})}\BibitemShut {NoStop}%
\bibitem [{\citenamefont {Duan}\ and\ \citenamefont
  {Guo}(1999)}]{duan1999suppressing}%
  \BibitemOpen
  \bibfield  {author} {\bibinfo {author} {\bibfnamefont {Lu-Ming}\ \bibnamefont
  {Duan}}\ and\ \bibinfo {author} {\bibfnamefont {Guang-Can}\ \bibnamefont
  {Guo}},\ }\bibfield  {title} {\enquote {\bibinfo {title} {Suppressing
  environmental noise in quantum computation through pulse control},}\ }\href
  {\doibase https://doi.org/10.1016/S0375-9601(99)00592-7} {\bibfield
  {journal} {\bibinfo  {journal} {Physics Letters A}\ }\textbf {\bibinfo
  {volume} {261}},\ \bibinfo {pages} {139 -- 144} (\bibinfo {year}
  {1999})}\BibitemShut {NoStop}%
\bibitem [{\citenamefont {Zanardi}(1999)}]{zanardi1999symmetrizing}%
  \BibitemOpen
  \bibfield  {author} {\bibinfo {author} {\bibfnamefont {Paolo}\ \bibnamefont
  {Zanardi}},\ }\bibfield  {title} {\enquote {\bibinfo {title} {Symmetrizing
  evolutions},}\ }\href {\doibase
  https://doi.org/10.1016/S0375-9601(99)00365-5} {\bibfield  {journal}
  {\bibinfo  {journal} {Physics Letters A}\ }\textbf {\bibinfo {volume}
  {258}},\ \bibinfo {pages} {77 -- 82} (\bibinfo {year} {1999})}\BibitemShut
  {NoStop}%
\bibitem [{\citenamefont {Viola}\ \emph {et~al.}(1999)\citenamefont {Viola},
  \citenamefont {Knill},\ and\ \citenamefont {Lloyd}}]{viola1999dynamical}%
  \BibitemOpen
  \bibfield  {author} {\bibinfo {author} {\bibfnamefont {Lorenza}\ \bibnamefont
  {Viola}}, \bibinfo {author} {\bibfnamefont {Emanuel}\ \bibnamefont {Knill}},
  \ and\ \bibinfo {author} {\bibfnamefont {Seth}\ \bibnamefont {Lloyd}},\
  }\bibfield  {title} {\enquote {\bibinfo {title} {Dynamical decoupling of open
  quantum systems},}\ }\href {\doibase 10.1103/PhysRevLett.82.2417} {\bibfield
  {journal} {\bibinfo  {journal} {Phys. Rev. Lett.}\ }\textbf {\bibinfo
  {volume} {82}},\ \bibinfo {pages} {2417--2421} (\bibinfo {year}
  {1999})}\BibitemShut {NoStop}%
\bibitem [{\citenamefont {Lidar}(2014)}]{lidar2014review}%
  \BibitemOpen
  \bibfield  {author} {\bibinfo {author} {\bibfnamefont {Daniel~A.}\
  \bibnamefont {Lidar}},\ }\enquote {\bibinfo {title} {Review of
  decoherence-free subspaces, noiseless subsystems, and dynamical
  decoupling},}\ in\ \href {\doibase 10.1002/9781118742631.ch11} {\emph
  {\bibinfo {booktitle} {Quantum Information and Computation for Chemistry}}}\
  (\bibinfo  {publisher} {John Wiley and Sons, Ltd},\ \bibinfo {year} {2014})\
  pp.\ \bibinfo {pages} {295--354}\BibitemShut {NoStop}%
\bibitem [{\citenamefont {Wallman}\ and\ \citenamefont
  {Emerson}(2016)}]{wallman2016noise}%
  \BibitemOpen
  \bibfield  {author} {\bibinfo {author} {\bibfnamefont {Joel~J}\ \bibnamefont
  {Wallman}}\ and\ \bibinfo {author} {\bibfnamefont {Joseph}\ \bibnamefont
  {Emerson}},\ }\bibfield  {title} {\enquote {\bibinfo {title} {Noise tailoring
  for scalable quantum computation via randomized compiling},}\ }\href@noop {}
  {\bibfield  {journal} {\bibinfo  {journal} {Physical Review A}\ }\textbf
  {\bibinfo {volume} {94}},\ \bibinfo {pages} {052325} (\bibinfo {year}
  {2016})}\BibitemShut {NoStop}%
\bibitem [{\citenamefont {Erhard}\ \emph {et~al.}(2019)\citenamefont {Erhard},
  \citenamefont {Wallman}, \citenamefont {Postler}, \citenamefont {Meth},
  \citenamefont {Stricker}, \citenamefont {Martinez}, \citenamefont
  {Schindler}, \citenamefont {Monz}, \citenamefont {Emerson},\ and\
  \citenamefont {Blatt}}]{erhard2019characterizing}%
  \BibitemOpen
  \bibfield  {author} {\bibinfo {author} {\bibfnamefont {Alexander}\
  \bibnamefont {Erhard}}, \bibinfo {author} {\bibfnamefont {Joel~J}\
  \bibnamefont {Wallman}}, \bibinfo {author} {\bibfnamefont {Lukas}\
  \bibnamefont {Postler}}, \bibinfo {author} {\bibfnamefont {Michael}\
  \bibnamefont {Meth}}, \bibinfo {author} {\bibfnamefont {Roman}\ \bibnamefont
  {Stricker}}, \bibinfo {author} {\bibfnamefont {Esteban~A}\ \bibnamefont
  {Martinez}}, \bibinfo {author} {\bibfnamefont {Philipp}\ \bibnamefont
  {Schindler}}, \bibinfo {author} {\bibfnamefont {Thomas}\ \bibnamefont
  {Monz}}, \bibinfo {author} {\bibfnamefont {Joseph}\ \bibnamefont {Emerson}},
  \ and\ \bibinfo {author} {\bibfnamefont {Rainer}\ \bibnamefont {Blatt}},\
  }\bibfield  {title} {\enquote {\bibinfo {title} {Characterizing large-scale
  quantum computers via cycle benchmarking},}\ }\href@noop {} {\bibfield
  {journal} {\bibinfo  {journal} {Nature communications}\ }\textbf {\bibinfo
  {volume} {10}},\ \bibinfo {pages} {1--7} (\bibinfo {year}
  {2019})}\BibitemShut {NoStop}%
\bibitem [{\citenamefont {et~al.}(2019)}]{qiskit}%
  \BibitemOpen
  \bibfield  {author} {\bibinfo {author} {\bibfnamefont {Gadi~Aleksandrowicz}\
  \bibnamefont {et~al.}},\ }\href {\doibase 10.5281/zenodo.2562110} {\enquote
  {\bibinfo {title} {Qiskit: An open-source framework for quantum computing},}\
  } (\bibinfo {year} {2019})\BibitemShut {NoStop}%
\bibitem [{\citenamefont {Strikis}\ \emph {et~al.}(2020)\citenamefont
  {Strikis}, \citenamefont {Qin}, \citenamefont {Chen}, \citenamefont
  {Benjamin},\ and\ \citenamefont {Li}}]{strikis2020learning}%
  \BibitemOpen
  \bibfield  {author} {\bibinfo {author} {\bibfnamefont {Armands}\ \bibnamefont
  {Strikis}}, \bibinfo {author} {\bibfnamefont {Dayue}\ \bibnamefont {Qin}},
  \bibinfo {author} {\bibfnamefont {Yanzhu}\ \bibnamefont {Chen}}, \bibinfo
  {author} {\bibfnamefont {Simon~C}\ \bibnamefont {Benjamin}}, \ and\ \bibinfo
  {author} {\bibfnamefont {Ying}\ \bibnamefont {Li}},\ }\bibfield  {title}
  {\enquote {\bibinfo {title} {Learning-based quantum error mitigation},}\
  }\href@noop {} {\bibfield  {journal} {\bibinfo  {journal} {arXiv preprint
  arXiv:2005.07601}\ } (\bibinfo {year} {2020})}\BibitemShut {NoStop}%
\bibitem [{\citenamefont {Czarnik}\ \emph {et~al.}(2020)\citenamefont
  {Czarnik}, \citenamefont {Arrasmith}, \citenamefont {Coles},\ and\
  \citenamefont {Cincio}}]{czarnik2020error}%
  \BibitemOpen
  \bibfield  {author} {\bibinfo {author} {\bibfnamefont {Piotr}\ \bibnamefont
  {Czarnik}}, \bibinfo {author} {\bibfnamefont {Andrew}\ \bibnamefont
  {Arrasmith}}, \bibinfo {author} {\bibfnamefont {Patrick~J.}\ \bibnamefont
  {Coles}}, \ and\ \bibinfo {author} {\bibfnamefont {Lukasz}\ \bibnamefont
  {Cincio}},\ }\href@noop {} {\enquote {\bibinfo {title} {Error mitigation with
  clifford quantum-circuit data},}\ } (\bibinfo {year} {2020}),\ \Eprint
  {http://arxiv.org/abs/2005.10189} {arXiv:2005.10189 [quant-ph]} \BibitemShut
  {NoStop}%
\end{thebibliography}%

\end{document}